# Deep UV laser induced periodic surface structures on silicon formed by self-organization of nanoparticles

Raul Zazo[1], Javier Solis[1], José A. Sanchez-Gil[2], Rocio Ariza[1], Rosalia Serna[1], and Jan Siegel[1]

1. Laser Processing Group, Instituto de Óptica, IO-CSIC, Serrano 121, 28006 Madrid, Spain
2. Instituto de Estructura de la Materia (IEM-CSIC), Consejo Superior de Investigaciones Científicas, Serrano 121, 28006 Madrid, Spain

Email: j.siegel@csic.es

**Abstract:**

We have investigated the formation of laser-induced periodic surface structures (LIPSS or "ripples") on silicon upon excitation with p-polarized excimer laser pulses in the deep ultraviolet region ($\lambda$ = 193 nm, 20 ns). Well-pronounced ripples with a period close to the laser wavelength were observed for pulse numbers N ≥ 100, and the ripple period increased with the angle of incidence. While these results seem to be qualitatively consistent with the standard Sipe-theory, we observed a fundamentally different ripple formation mechanism and ripple morphology. At low pulse numbers, isolated nanoparticles with a size of a few tens of nanometers are observed at the silicon surface, which then start to agglomerate in 2D and self-organize to form ripples with a very shallow modulation depth as the pulse number increases. Employing a recently developed plasmonic model based on the propagation of a surface plasmon polariton on a rough surface, we demonstrate excellent quantitative agreement of the evolution of the ripple period with incidence angle. Finally, we show that surface regions exposed to lower laser fluence feature micro- and nanopores, which give rise to pronounced photoluminescence (PL) emission in the visible spectral region, as opposed to the nanoparticle-based ripples not showing PL.

## 1. Introduction

Since their discovery in 1965, laser-induced periodic surface structures (LIPSS) have attracted the attention of the scientific community due to their formation mechanisms based on self-organization and their potential for numerous technological applications [1]. Briefly, LIPSS are formed when a material is exposed to multiple short or ultrashort laser pulses at energies above the modification threshold [2–4]. The underlying mechanism that is generally held responsible for LIPSS formation is related to the presence of a transient periodic intensity modulation at the surface, formed by the interference of incident laser light with a wave propagating at the surface, and which is eventually imprinted into the material via local ablation [5]. LIPSS have been reported in metals [6–8], semiconductors [9–11] and dielectrics [12–16], and feature a variety of different sizes, shapes and orientations that are determined by laser irradiation parameters (fluence, wavelength, pulse number and polarization being the most important ones) together with several material properties. The broad range of different structures that can be fabricated by simple laser processing, exploiting self-interference and self-organization of matter within a typical laser spot size of tens to hundreds of micrometers, has triggered intense research over the past decades [3,17–21].

The arguably most widely investigated LIPSS type are so-called low spatial frequency LIPSS (LSFL or "ripples") which have mainly been observed in the ablation regime, leading to a considerable periodic modulation of the surface topography [2,4]. A few works report on ripple formation mechanisms other than local ablation, namely crystalline-amorphous phase change [10,22–24], surface oxidation [8], as well as self-arrangement of preexisting nanoparticles embedded in thin films [25,26]. Concerning the lasers employed, early works have investigated ripple formation in semiconductors upon nanosecond (ns) laser excitation in the visible and near infrared (NIR) region [2], although interest has shifted soon towards LIPSS formed upon NIR femtosecond laser excitation. Although the ripple period is known to scale with the laser wavelength, only few works report on ultraviolet (UV) laser induced formation of LIPSS [27,28]. One possible reason for the small number of studies at this wavelength range is that most of the available lasers operating in the UV region emit non-polarized light, which prevents the fabrication of ripple-type LIPSS.

In our study, we have used a non-polarized nanosecond excimer laser operating at 193 nm (deep UV) and developed a simple irradiation setup that allows laser excitation of silicon with p-polarized pulses at variable angles of incidence. These conditions were found to trigger the formation of a different type of ripples, based on the formation, agglomeration and self-organization of laser-generated nanoparticles. To the best of our knowledge, such type of ripple formation has been reported only in one occasion in the UV (248 nm excitation) [28]. By systematically investigating the influence of the pulse number and angle of incidence on the surface morphology and topography, and applying a recently developed roughness-mediated plasmonic model we are able to both, quantitatively describe the evolution of the period with incidence angle and monitor the ripple formation process.

## 2. Experimental

Fig. 1 shows the experimental setup for ripple fabrication using a deep UV excimer laser. While the concept of the layout is similar to the one used in our previous work to yield s-polarized light [29], the present configuration has been adapted to yield p-polarization. Briefly, we used an ArF excimer laser delivering non-polarized laser pulses at $\lambda$ = 193 nm and t = 20 ns pulse duration. The pulse energy could be continuously adjusted by rotating a coated fused silica window inserted into the beam path, whose transmission depended strongly on the angle of incidence. The laser was operated typically at repetition rates between 1 and 10 Hz. The excimer laser beam was polarized by inserting a fused silica prism at Brewster angle (57º at 193 nm) in the beam path and selecting the reflected light for irradiation, following the concept used in Refs [30,31]. This configuration eliminates the vertical (s-) polarization component and provides a strongly horizontally (p-) polarized beam for irradiation, which was focused by a fused silica biconvex lens (focal length f = 90 mm) onto the sample surface.

The sample was mounted on a manually controlled x-y-z stage with additional control of the angle of incidence $\theta$. The intensity distribution incident on the sample was designed to be top hat-like by using an imaging setup, i.e. by means of the focusing lens projecting a 5 mm diameter circular aperture (Aperture 1 in Fig. 1) onto the sample surface, which was positioned in the image plane. Absolute pulse energy measurements were performed using a calibrated pyroelectric energy detector. The corresponding laser fluences were calculated by dividing the laser pulse energy by the area of the image of Aperture 1, whose average diameter at normal incidence was determined experimentally through ablation measurements of the slightly elliptical spots to be D = 250 μm. Since the pulse energy deposited in the sample depends on the angle of incidence due to the angle-dependent Fresnel reflection, we adjusted the pulse energy for each angle such that the absorbed fluence was constant, rather than the incident fluence. This was achieved by ensuring that the same vertical spot diameter and optical contrast upon inspection with an optical microscope for N = 100 pulse was maintained for each angle. For θ = 0º, this corresponds to F = 1.30 J/cm². In-situ monitoring of the laser-induced changes was performed by means of a home-build imaging setup, based on a CCD camera with a macro lens together with a LED for illumination. The experimental setup used for ripple fabrication with femtosecond laser pulses (used for obtaining the results shown in Fig. 2b) can be found in Ref [32]. Briefly, an Yb fiber laser provided 350 fs pulses at a central wavelength of 1030 nm and a repetition rate of 500 kHz, which were sent through scan head with galvanometer-driven mirrors followed by an F-Theta lens that focused the beam down to a spot of 48 μm in diameter. The spot was scanned over the sample surface at a constant speed v = 200 mm/s, which corresponds to an effective pulse number of $N_{eff}$=120.

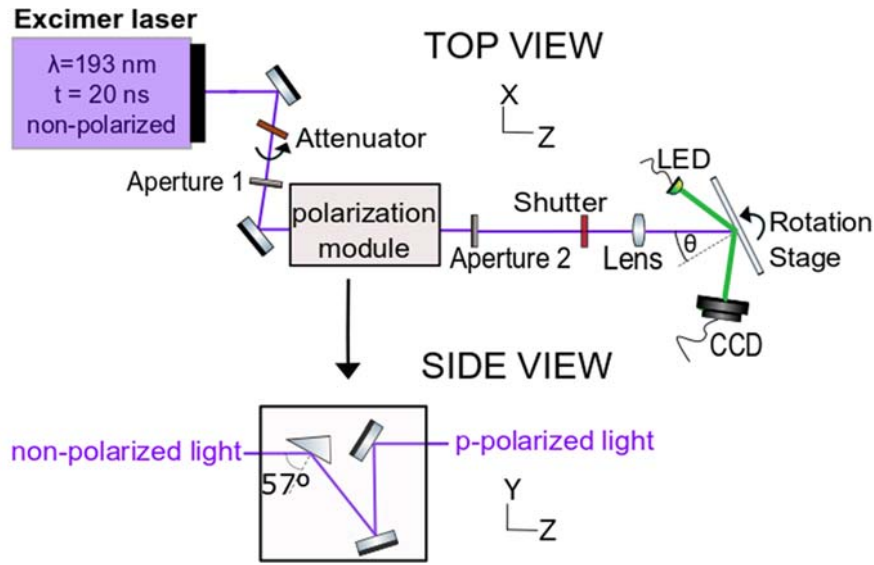

*Fig. 1: Excimer laser-based irradiation setup used to generate nanoparticle-based LIPSS.*

The samples used were commercial <100> oriented crystalline silicon wafers with p-doping (boron, resistivity 1–5 Ω cm), cleaned before irradiation with acetone in an ultrasonic bath. After irradiation, the samples were inspected by a Nikon Eclipse Ti inverted microscope with a 100x dry objective lens (N.A. = 0.9) and LED illumination at 460 nm, yielding a lateral resolution $R_{xy}$ < 300 nm. The surface morphology and topography has been characterized by a scanning electron microscope and an atomic force microscope operating in tapping mode, respectively. Both instruments were made accessible through the NFFA-Europe program. The determination of the ripple period Λ from SEM images was performed via two-dimensional Fast Fourier Transform (2D FFT), whereas horizontal profiles, averaging over 20 rows, were extracted from AFM maps, which then were analyzed via 1D FFT.

Photoluminescence (PL) measurements were performed using continuous wave laser excitation at 355 nm with a nominal power of 4 mW and a lens with a focal distance of f = 150 nm, incident at an angle β = 55º. The luminescence emitted by the samples was collected in the normal direction of the sample surface with a x10, NA = 0.26 objective lens and imaged by means of a tube lens with f = 100 mm onto the 2 mm slit of a Czerny-Turner type monochromator and detected by a photomultiplier. The signal was amplified using a standard lock-in technique employing an optical chopper and collected by a PC. All PL spectra were recorded at room temperature. Alternatively, to point excitation, the setup could be switched to wide field excitation and the PL was imaged onto a CCD camera, which allowed a precise sample positioning for location specific PL-measurements.

## 3. Results and discussion

*3.1 Ripple period dependence on laser wavelength λ*

Figure 2 shows SEM images of low spatial frequency LIPSS (LSFL or "ripples") formed upon irradiation with two different laser systems at normal incidence ($\theta = 0°$). Using N=100 linearly polarized femtosecond (fs) laser pulses in the NIR ($\lambda = 1030$ nm and $\tau = 340$ fs) at fluences near the single pulse ablation threshold, well pronounced ripples with orthogonal orientation with respect to the polarization direction are obtained (cf. Fig. 2b). This type of ripples has been widely studied and can easily be extended over large areas by beam scanning [33], their period being slightly below the laser wavelength ($\Lambda_{NIR-fs} = 840$ nm; $\lambda/\Lambda(NIR-fs) = 0.82$). Their characteristic morphology clearly shows that the material has been melted and resolidified, and that ablation has occurred in the trenches, giving rise to a pronounced modulation of the surface topography. The mechanisms that lead to this type of structure include the incoupling of SPPs from the incident light with help from a non-negligible surface roughness, propagation of the SPP at the air/material interface, interference with the incident laser pulse, localized melting and/or ablation at the interference maxima, in some cases melt flow/hydrodynamics, and finally resolidification, which can be either in the crystalline or amorphous phase [10,34,35].

For comparison, the ripples structures we report here for irradiation with nanosecond laser pulses in the deep UV ($\lambda = 193$ nm and $\tau = 20$ ns) are shown in Fig. 2a. Despite the very different laser parameters in terms of wavelength and pulse duration, the ripples appear to be similar, namely well aligned and perpendicular to the polarization orientation, as well as their period close to the laser wavelength in this case ($\Lambda_{UV-ns} = 164$ nm; $\lambda_{UV}/\Lambda(UV-ns) = 0.85$). However, as shown and discussed in the following sections, they feature remarkable differences both in terms of nanoscale morphology and formation mechanisms.

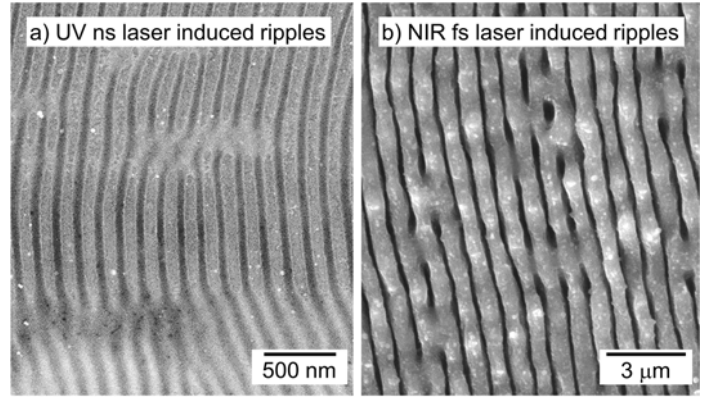

*Fig. 2: Comparison of ripples on silicon, fabricated with different laser wavelengths λ and pulse durations t, but using in both cases normal incidence, horizontal polarization (p-polarized) and a similar number of pulses N a) λ = 193 nm, t = 20 ns, N = 100. b) λ = 1030 nm, t = 340 fs, N = 120. Please note the different scale bars.*

*3.2 Ripple period dependence on angle of incidence θ*

Concerning the period of ripple-type LIPSS, it is well known that it can be changed up to a certain extent by varying the angle of incidence [22,35,36]. For metals, taking into account the propagation of surface plasmon polaritons (SPP) at the surface, the following expression is obtained [37,38]:

$$\Lambda^{\pm} = \frac{\lambda}{\text{Re}[\eta] \mp \sin\theta} \quad (1)$$

where θ is the angle of incidence with respect to the surface normal, λ the laser wavelength and Re[η] is the real part of the complex refractive index η of the air-metal interface, with

$$\eta = k_{SPP}^0 \cdot \frac{\lambda}{2\pi} = \left[\frac{\epsilon_{air}\epsilon_{metal}}{(\epsilon_{air}+\epsilon_{metal})}\right]^{1/2} \quad (2)$$

$\epsilon_{air}$ and $\epsilon_{metal}$ are the complex dielectric functions of air and metal, respectively, and $k_{SPP}^0$ represents the SPP wavevector. Notably, this model predicts two ripple periods, $\Lambda^+$ and $\Lambda^-$, when irradiating at oblique incidence, which has indeed been observed experimentally in a variety of materials. While this model is in principle limited to metals that dispose of free electrons to support SPP propagation, it has been extended to semiconductors due to the fact that these also readily form ripples [22,39]. In order to explain this behavior, it is often argued that the dielectric function $\epsilon_{semiconductor,excited}$ of the transiently laser-excited material needs to be taken into account for LIPSS formation, rather than the steady-state dielectric function [35,40]. Unfortunately, a measurement of this transient property is difficult. Another, much simpler, way to obtain a realistic value for η is to use the dielectric function of the molten phase $\epsilon_{semiconductor,melt}$, of the semiconductor, which is metal-like and thus supports SPP propagation [41]. Using femtosecond microscopy to study the dynamics of phase-change LIPSS formation in silicon, it has been shown that the lifetime of the molten phase during each irradiation pulse is approximately 3 nanoseconds, which supports the idea of using the properties of the liquid phase [35]. Table 1 displays the optical properties of the silicon in the crystalline and molten phases at λ = 193 nm used in the present study.

| Material | $\varepsilon_1$ | $\varepsilon_2$ | Reference | Re[η] (eq. (2)) |
|---|---|---|---|---|
| Air | 1 | 0 | | |
| c-Si | -7.4422 | 5.8762 | [42] | 1.04 |
| m-Si | -15.0 | 3.0 | [43] (extrapolated) | 1.03 |

*Table. 1: Real part $\varepsilon_1$ and imaginary part $\varepsilon_2$ of the dielectric function of air, crystalline Si (c-Si) and molten Si (m-Si), accompanied by the references the data were taken from and the obtained value for the real part of the complex refractive index η of the air-metal interface, at which SPP propagation takes place.*

We have experimentally investigated the angle of incidence dependence of the ripples formed upon deep UV ns laser irradiation at constant pulse number N = 100. Figs. 3(a-c) show high magnification SEM images for the case of θ = 0º, 30º, 50º, respectively. In all cases, well-pronounced ripples aligned perpendicular to the laser polarization were observed, with the ripple widths and periods $\Lambda^+$ clearly increasing with the angle of incidence. At these high magnification images, a pronounced corrugation at the ripple border can be seen, whereas the inner region of a ripple is relatively featureless, even for the widest ripples shown (θ = 50º).

It is worth mentioning that for some angles we could observe formation of the shorter periods $\Lambda^-$ predicted by eq. (1). As shown for the case of θ = 20º in Fig. 3d), these were found typically at the border of the irradiated region, corresponding to a lower local fluence. It should be mentioned, though, that such a spatial-fluence separation is not intrinsic to the general formation mechanism of ripples. In fact, other works report the spatial-fluence coexistence of $\Lambda^+$ and $\Lambda^-$ [44] ripples, and preferential selection of one ripple period type can even be achieved by an adequate choice of the processing parameters [35].

The evolution of the ripple period $\Lambda^+$ as a function of incidence angle is plotted in Fig. 3(e), featuring a monotonous increase. The plot also includes the theoretical curve calculated according to eqs. 1 and 2, using the dielectric function of molten silicon ($\varepsilon$ = -15.0 + 3.0i; cf. table 1), which yields a poor match. Using the dielectric function of crystalline silicon provides an equally poor match (curve not displayed), which underlines the need of a modified model to describe the experimental results. The red curve, included in Fig. 3(e), describes reasonably well the behavior of the experimental data and has been obtained by using Re[$\eta$] as a fitting parameter, yielding a value of Re[$\eta$] = 1.12.

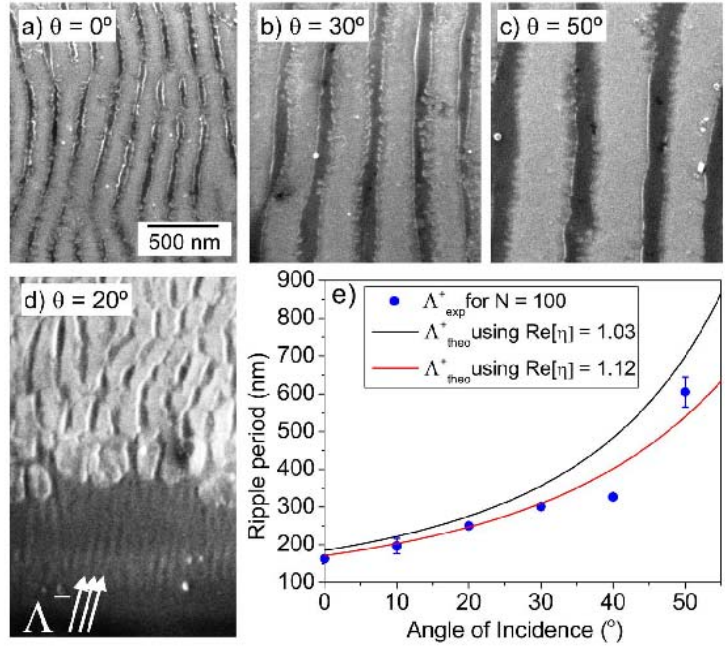

*Fig. 3: Evolution of the ripple morphology and period $\Lambda^+$ upon deep UV irradiation (193 nm and 20 ns, N = 100) of silicon at different angles of incidence $\theta$. a) – d) SEM images at selected angles. The scale bar displayed in a) applies to all images. For the particular case of d) $\theta = 20°$, ripples with an additional period $\Lambda^-$ can be seen in the lower part of the frame. e) Evolution of the experimentally determined ripple period $\Lambda^+_{exp}$ with angle of incidence, compared to calculated values $\Lambda^+_{theo}$ according to eq. (1), using different values for Re[$\eta$] (see text for details).*

The motivation of such high value lies in the influence of the significant modification of the SPP wave vector by the presence of the surface roughness induced during multiple laser pulse irradiation. Very recently, we have introduced a model that takes into account the influence of the specific roughness properties on the SPP wave vector as $k^0_{SPP} = k^{(0)}_{SPP} + \Delta k_{SPP}$, described in detail in Ref. [45], which has a direct influence on Re[$\eta$] in eq. (2). This model is able to correctly describe the experimentally observed ripple periods in copper [45] and which we could extend successfully to absorbing metals such as steel [44]. A related model has been proposed by Ionin et al., who considered scattering of SPPs on surface relief features with a certain relief height, which predicts a corresponding shift of the SPPs' dispersion curve [46]. For the case of silicon, they reported a qualitative agreement in terms of period reduction as a function of relief height, although no quantitative match was obtained. We will show in the next section by means of AFM data that the surface roughness is indeed significantly modified in our case, which has a direct impact on the SPP wavevector and leads to a decrease of the ripple period.

*3.3 Evolution of the topography, morphology, and ripple period with pulse number*

Figs. 4a)-c) display topography maps of ripple structures formed after irradiation with different pulse numbers at θ = 50º. While only weak order can be appreciated at N = 10 pulses, periodic ripples emerge at N = 20 and are fully developed at N = 100. Interestingly, a substructure of the weakly ordered ripples can be appreciated at N = 10, consisting of agglomerated nanoparticles with an average lateral diameter of around d = 80 nm, possibly overestimated by the large step size of the AFM measurement (10 nm). For comparison, the height modulation of the structures for the different pulse numbers is displayed in Fig. 4d). At N = 10, the modulation depth is Δh < 3nm, which is much less than the apparent lateral extension. This demonstrates that the nanoparticle agglomerates are rather planar at this pulse numbers, with an approximate aspect ratio d/Δh > 26.

At N = 20, the nanoparticle substructure begins to fade, accompanied by a slight increase in modulation depth, reaching Δh = 5nm. Finally, at N = 100, smooth and dense ripples can be observed with a much higher modulation of Δh = 15nm. It should be emphasized at this point that this height modulation is extremely small compared to standard LIPSS in Si reported upon irradiation with NIR fs laser pulses. For instance, for the specific result shown in Fig. 2(b) obtained for the same pulse number but with fs laser irradiation in the NIR, we have measured $\Delta h_{NIR,fs}$ > 700 nm (results not shown), which is a factor 50 higher.

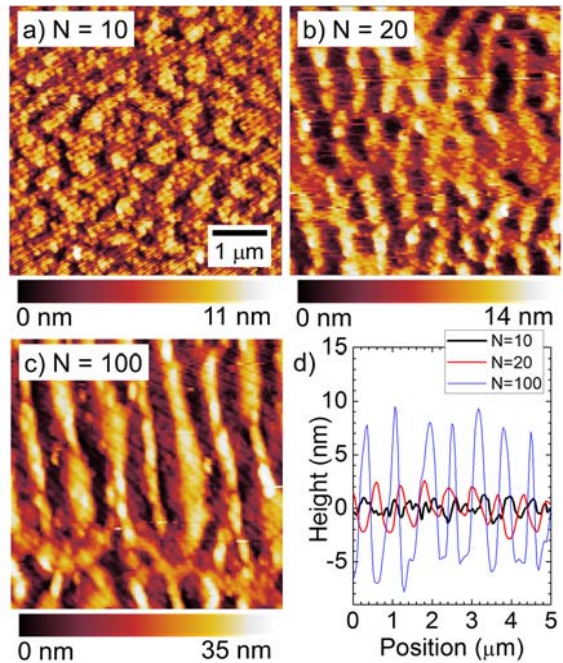

*Fig. 4: Evolution of the surface topography upon deep UV irradiation (193 nm and 20 ns, θ = 50º) of silicon at different pulse numbers. a) – c) AFM images at selected pulse numbers N. The scale bar displayed in a) applies to all images. d) Horizontal height profiles through the center of the images, averaging over 16 rows (160 nm).*

Concerning the influence of the surface roughness on the SPP wavevector $k_{SPP}^0$, and thus on Re[η] in eq. (2) as discussed in section 3.2, we have used the experimentally determined surface roughness parameters RMS height deviation δ and correlation length σ to calculate the expected ripple period using the model introduced in Ref. [45]. The expression used there has been obtained through surface roughness perturbation theory [47,48], which also assumes that the imaginary part of the dielectric function of the material investigated is much smaller than the real part [48], $|\epsilon_r| \gg |\epsilon_i|$, which is indeed the case for the molten silicon (cf. table 1). One of the constraints of the model is, though, that the roughness parameters used necessarily need to be determined from a surface exposed to a pulse number immediately before the appearance of ripples, since a determination of the correlation length $\sigma$ is not possible for periodic structures. Applied to our case, this corresponds to N = 10, which yields -upon analysis of the corresponding AFM map displayed in Fig. 4a)- values of $\delta$ = 1.1 nm and $\sigma$ = 90 nm. Using these values as input for our model yielded again a poor fit of the predicted period with the experimental data. However, upon increasing the $\delta$–value up to $\delta$ = 8 nm while maintaining $\sigma$ = 90 nm, our model yields exactly the same curve as displayed in Fig. 3e obtained by fitting, showing a good match with the experimental data. We motivate the necessary strong increase of the RMS height deviation $\delta$ as a suitable input value for our model by the fact that a considerable increase in modulation height is observed at higher pulse numbers (e.g. Δh = 15nm at N = 100, see Fig. 4d).

In order to further investigate the presence of nanoparticles and their contribution to the ripple formation we have performed an additional irradiation study as a function of pulse number, but this time at θ = 0°, in order to obtain the shortest period (cf. eq. 1). Fig. 5 shows representative SEM images at different N-values. The image at N = 20 (Fig. 5a) unambiguously demonstrates that the initial state of ripple formation consists in the agglomeration of nanoparticles. The inset shows a magnification of a small region of the same image, featuring few isolated nanoparticles, which allows an estimation of their size, approximately d ≈ 20 nm. This observation is consistent with the results reported by Fowlkes et al. upon irradiation at λ = 248 nm in low pressure He atmosphere [49], showing the presence of nanoparticles of a similar size, which could be arranged into LIPSS like structures by a two-step irradiation strategy.

At higher pulse number (Figs. 5b-d), the nanoparticles further agglomerate and form vertical narrow lines (N = 50), which widen and smoothen to form ripples (N = 100, 200). In order to determine if the considerable change in ripple morphology influences the ripple period, we have determined the latter via FFT of the SEM images and plotted the obtained period as a function of pulse number in Fig. 5e). The results clearly show that at low pulse numbers (N = 20), and thus low roughness, the ripple period is significantly higher than at N = 100, which corresponds to the value plotted in Fig. 3. Moreover, this behavior is fully consistent with the model introduced, which predicts a shortening of the period as the surface roughness increases.

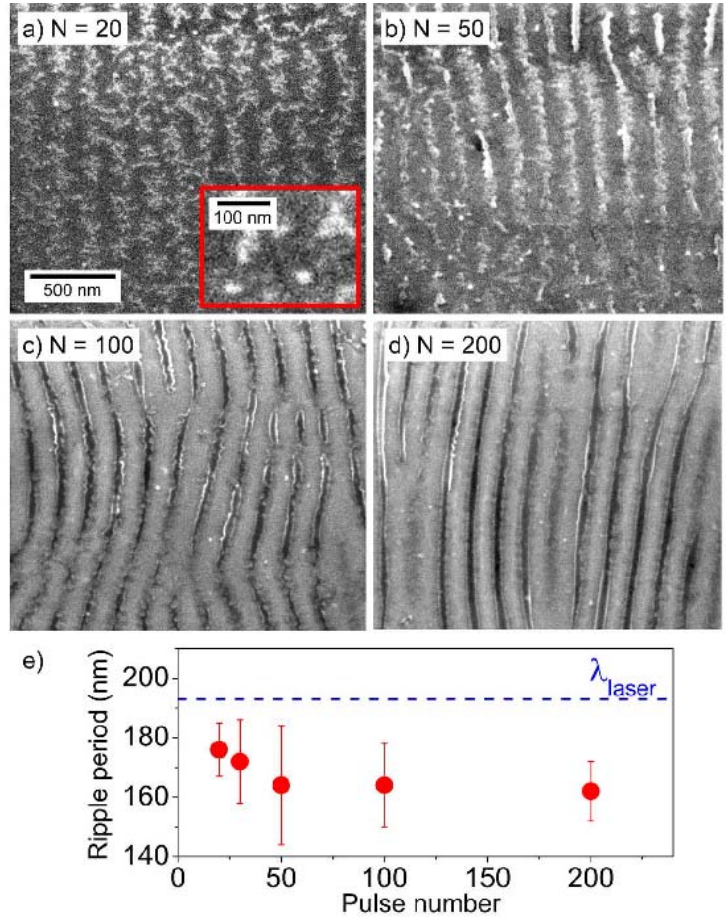

*Fig. 5: Evolution of the surface morphology upon deep UV irradiation (193 nm and 20 ns, θ = 0°) of silicon at different pulse numbers. a) – d) SEM images at selected pulse numbers N. The scale bar displayed in a) applies to all images, while the inset included (with its own scale bar) corresponds to a magnification of a small region in a). e) Ripple period as a function of pulse number determined from SEM images recorded at a wide range of angles.*

### *3.4 Photoluminescence of excimer laser irradiated silicon*

Since silicon is an indirect bandgap semiconductor, light emission from bulk is virtually impossible. Yet, being a key material in electronics and optical communications, strategies for turning silicon into a material that allows light emission are of great interest. In this context, porous silicon has been identified as a suitable candidate to provide efficient luminescence due to the presence of crystalline Si nanostructures [50]. Moreover, irradiation studies of silicon using femtosecond pulses have reported visible luminescence in form of a Gaussian-like spectrum covering the 450 nm – 800 nm range [51]. The authors attributed these results to the formation of hierarchical structures consisting in micro-sized cones covered by dendritric Si nanostructures, at which oxygen from the surrounding air was incorporated into the silicon lattice.

We have investigated if the deep-UV laser irradiated nanostructures fabricated in the previous sections might show similar photoluminescence (PL) properties. We have found that no significant PL is observed in the central regions that contain the nanoparticle-assembled ripples. However, we observe a strong PL emission at the border region of the spot, a region that has been exposed to a lower laser fluence. As can be seen in Fig. 6a), the morphology at this region is very rough, featuring micro- and nanopores, as opposed to the very smooth central region that features the ripples. Interestingly, the PL spectrum shown in Fig. 6b) also extends throughout the visible range (400 – 700 nm), similar to the one reported for fs laser irradiation [51]. Moreover, also in our case the PL intensity is found to depend on the pulse number, reaching a maximum for N = 500. Further studies to investigate the particular nanostructures and compositional changes at this border region that give rise to this pronounced PL emission are on the way.

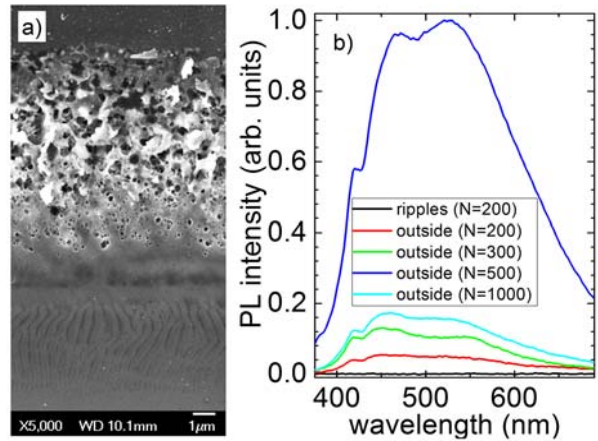

*Fig. 6: Spatially resolved photoluminescence study of Si laser-irradiated at 193 nm, 20 ns, θ = 40º Si samples. a) SEM image of the transition between the region featuring ripples towards the spot center (bottom) and the outer region featuring micro- and nanoporous structures (upper part). b) Photoluminescence spectra recorded in the different regions for different pulse numbers upon excitation at 355 nm.*

**Conclusions:**

Using linear polarized nanosecond laser pulses at 193 nm (deep UV) we observe a seldomly reported form of laser induced periodic surface structures (LIPSS) on silicon. As opposed to conventional LIPSS, which are mainly formed by localized ablation at interference maxima, the LIPSS presented here form upon laser-induced generation, agglomeration and self-organization of nanoparticles. As a consequence, their morphology strongly depends on the pulse number, and a minimum of N = 20 pulses is required to start forming periodic structures composed of in-plane near-coalescence nanoparticles with a diameter of around 20 nm, which progressively evolve with pulse number into smooth and flat ripples (height less than 15 nm) composed of densely-packed nanoparticles. The period of these structures is found to be close to the laser wavelength used (for normal laser incidence), and decreases slightly when increasing the pulse number. At oblique angles of incidence, the period increases as expected according to the simple plasmonic or scatter model for such structures, yet the absolute period values are smaller than predicted. By considering the specific surface roughness, our recently developed roughness-mediated plasmonic model is able to describe reasonably well the period evolution with angle, although the roughness parameters used correspond to the final structures formed rather than to the roughness at the threshold of ripple formation. The nanoparticle-nature of these ripple structure could be of interest for potential applications, such as sensing. While the LIPSS structures themselves do not exhibit photoluminescence, the border region of the irradiated area composed of micro- and nanopores shows a pronounced photoluminescence emission in the visible spectral region, which might be of interest for photonics applications.

**Acknowledgements**

JSi, JSo, and RS acknowledge financial support through the national research grants UDiSON (TEC2017-82464-R) and SENSIL (RTI2018-096498-B-I00) from the Spanish Research Agency (AEI, Ministry of Research and Innovation) and the European Regional Development Fund (ERDF), as well as the Consejo Superior de Investigaciones Científicas for the intramurales project (201850E057) and PICS project SAMUL (2018FR0026) jointly with the CNRS, France. The authors also acknowledge pre-doctoral grants to RZa and RAr awarded by the regional government of Madrid and financed by the European Social Fund (ESF) and the national "Iniciativa de Empleo Juvenil" (YEI). The authors acknowledge the NFFA EUROPE platform for financial support and access with technician support to SEM and AFM equipment studies at the facilities of FORTH, Heraklion-Greece within project ID 507. JASG acknowledges the Spanish MCIU/AEI/FEDER for financial support through the grants MELODIA (PGC2018-095777-B-C21) and NANOTOPO (FIS2017-91413-EXP).